\documentclass[12pt,epsfig]{article}
\begin{document}
\setlength{\oddsidemargin}{0in}
\setlength{\textwidth}{6.0in}
\setlength{\textheight}{8.0in}

\newcommand{\nc}{\newcommand}
\newcommand{\beq}{\begin{equation}}
\newcommand{\eeq}{\end{equation}}
\newcommand{\bear}{\begin{eqnarray}}
\newcommand{\ear}{\end{eqnarray}}
\newcommand{\bi}{\bibitem}
\newcommand{\rar}{\rightarrow}
\newcommand{\lar}{\leftarrow}
\newcommand{\lrar}{\leftrightarrow}

\begin{center}
\vglue .06in

{\large
{\bf {Generation of Magnetic Fields in Cosmology}}
\bigskip
\\{A.D. Dolgov}
 \\[.5in]
{\it
{INFN, sezione di Ferrara\\ Via Paradiso, 12 - 44100 Ferrara,
Italy \\
and\\
ITEP, B. Cheremushkinskaya 25, Moscow, 117259, Russia
}}}
\end{center}

\begin{abstract}
Mechanisms of generation of magnetic fields in the early universe 
which could seed the present-day large scale galactic magnetic fields, 
are briefly reviewed. Three possible ways to create large scale 
magnetic fields are discussed: breaking of conformal invariance of 
electromagnetic interactions and inflationary stretching of the field
wave length, first order cosmological phase transitions, and chaotic 
electric currents generated by turbulent flows in the primeval plasma.
\end{abstract}

\section{Introduction}\label{sec:intr}

Astronomical observations show that there exist magnetic fields in 
galaxies with the field strength about 1 micro-gauss, coherent on the 
whole galactic size; for recent detailed reviews 
see~\cite{kronberg94,grasso01}.
Though magnetic fields of individual stellar bodies can be and are much
larger, an existence of coherent fields on the scales about 1 kpc or 
bigger presents one of the most profound mysteries of modern cosmology.
There are many simple and realistic mechanisms of creation of magnetic
fields in the early universe on small scales, however it is difficult 
to make them operate at galactic size because the latter is much larger
than horizon at the epoch of field generation. Another problem is that
the background cosmological model is homogeneous and isotropic and,
though some primordial density perturbations should exist, they are 
assumed to be generated at inflationary stage and are normally of scalar 
(or tensor) type, i.e. vorticity free. This is also unfavorable with
respect to formation of vector fields on macroscopic scales. Rather large
vorticity perturbations might be generated in the course of possible first
order electroweak or QCD phase transitions when the bubbles of new phase
were formed in the old one and a strong, though small scale, chaoticity is 
excited in the plasma. It is worth mentioning in this connection that
classical solutions of equations of motion of non-abelian gauge fields
and the Einstein equations demonstrate chaotic behavior, as was shown in 
a series of papers by S. Matinyan and collaborators~\cite{matinyan00}
(and references therein). It is an interesting question if this chaoticity 
may be related to magnetic field generation in cosmology.

Roughly speaking there are three possible mechanisms of magnetic field
creation in the early universe discussed in the literature:

1. Breaking of conformal invariance of electromagnetic interaction at 
inflationary stage. The latter could be realized either through new
non-minimal (and possibly non gauge invariant) coupling of 
electromagnetic field to curvature~\cite{turner88}, or in dilaton
electrodynamics~\cite{ratra92}, or by the well known conformal anomaly 
in the trace of the stress tensor induced by quantum corrections
to Maxwell electrodynamics~\cite{dolgov93}.

2. First order phase transitions in the early universe~\cite{hogan83}
producing bubbles of new phase inside the old one. A different mechanism
but also related to phase transitions is connected with topological
defects, in particular, cosmic strings~\cite{vachaspati91}.

3. Creation of stochastic inhomogeneities in cosmological charge 
asymmetry, either electric~\cite{dolgov-s93}, or e.g. 
leptonic~\cite{dolgov01} at 
large scales which produce turbulent electric currents and, in turn, 
magnetic fields.

In what follows I will briefly describe these three mechanisms. The 
literature on the subject is very rich and it is impossible to discuss 
all the relevant papers in this short contribution, so I will quote 
mostly only the original works where the idea was formulated and 
the latest ones where the proper list of references can be found.

\section{Breaking of Conformal Invariance of Electrodynamics}

It was established long ago that conformally flat gravitational field 
(in particular the usual cosmological Friedman-Robertson-Walker 
background) does not produce massless particles if the underlying 
theory is conformally invariant~\cite{parker68}. However, both 
gravitons~\cite{grishchuk75} and minimally coupled to gravity massless 
scalars are not invariant and this gives rise to creation of density
perturbations~\cite{denpert} and gravitational waves~\cite{rubakov82}
of very large wave length, if the are produced 
at inflationary stage. On the other hand,
classical electrodynamics is known to be conformally invariant, so that 
photons should not be produced in cosmological background.
However if one introduces a new type of interaction into the Maxwell 
Lagrangian, then the invariance may be broken and long wave electromagnetic 
fields could be generated at inflation by the same mechanism as 
gravitational waves. The former could be quasistatic and serve as seeds 
of coherent galactic magnetic fields.

A model of this kind was proposed in ref.~\cite{turner88}, where a 
non-minimal coupling of electromagnetic field to gravity was considered:
\bear
{\cal L} = -{1\over 4} F_{\mu\nu} F^{\mu\nu} + C_1 R A_\mu A^\mu +
C_2 m^{-2} R_{\mu\nu} F^{\mu\alpha} F^{\nu}_{\alpha} +...    
\label{nonmin}
\ear
Here $F_{\mu\nu}$ is the electromagnetic field tensor, $R_{\mu\nu}$ is 
the Ricci tensor, $R$ is the curvature scalar, $C_j$ are dimensionless
constants, and $m$ is another constant with dimension of mass. The first
term is the usual free Maxwell Lagrangian, while the others represent new
hypothetical interaction which breaks conformal invariance. The second
term breaks also gauge invariance and would give a non-zero mass to 
photons in space-time with non-vanishing curvature.  

Another example of non-standard electrodynamics was discussed in 
ref.~\cite{ratra92} in the dilaton electrodynamics where the free
Maxwell Lagrangian was modified by the coupling to dilaton field
$S$ in the following way:
\bear
{\cal L} = -{1\over 4}\, \exp \left(S/\eta \right)\,\, 
F_{\mu\nu} F^{\mu\nu},
\label{dilaton}
\ear
where $\eta$ is a constant parameter with dimension of mass.
The dilaton model of magnetic field generation finds its natural 
realization in string cosmology~\cite{gasperini95}; for a review of
the latter see e.g.~\cite{gasperini00}. A related idea of magneto-genesis
from time variation of gauge couplings was discussed recently in 
ref.~\cite{giovannini01}. In a special case of dilaton theory this 
model corresponds to time variation of the dilaton field $S(t)$.

On the other hand, the usual quantum electrodynamics is not conformally
invariant because of quantum corrections. In particular the trace
of energy-momentum tensor which should be zero in conformally invariant
theory, becomes non-vanishing due to triangle diagram with the electron 
loop~\cite{chanowitz73}. This quantum anomaly results in the following
modification of the Maxwell equations~\cite{dolgov81}:
\bear
\partial_\mu F^\mu _\nu + \kappa {\partial_\mu a \over a} F^\mu_\nu =0
\label{anomeq}
\ear
where $a=a(\tau ) $ is the cosmological scale factor, $\tau$ is the 
conformal time,
the metric has the form $ds^2 =a^2(\tau )(d\tau^2 -dr^2 )$, and the 
contraction of the indices is made with the metric tensor of the flat 
space-time. The numerical coefficient
$\kappa$ in $SU(N)$-gauge theory with $N_f$ charged fermions is
equal to $\kappa = {\alpha/\pi}\left( {11N/ 3} -{2N_f/ 3}\right)$.
Here $\alpha $ is the fine structure constant which is to be taken at the
momentum transfer $p$ equal to the Hubble parameter during inflation,
$p=H$. In the asymptotically free theory one would expect
$\alpha \approx 0.02$.

The additional anomalous term in eq~(\ref{anomeq}) gives rise to the 
production of photons by conformally flat gravitational field
and, thanks to inflation, to generation of large scale magnetic 
fields~\cite{dolgov93}. The magnitude of the effect is too small in the 
case of the contribution of one electron loop but in theories with many
charged particles, as e.g. in grand unified theories, the effect may be 
significant. The masses of these particles should be small in comparison
with the Hubble parameter during inflation. A detailed analysis of 
magnetic field generation due to quantum anomaly can be found in the
recent work~\cite{prokopec01}. 

A way to break conformal invariance of gauge fields during inflation
due to coupling to a light ($m<H$) charged scalar field, $\phi$, was  
suggested in ref.~\cite{turner88} and studied in a special model
in the papers~\cite{calzetta98}. The authors argued that stochastic
electric currents could be generated during inflation due to production 
of charged scalar particles $\phi^\pm$ by the inflaton field. 
However,  this scenario was criticized in refs.~\cite{giovannini00}, 
where it was shown that dissipative effects induced by the plasma 
conductivity, which were disregarded in the above quoted papers, 
would strongly diminish the currents making them too weak to 
seed the the galactic magnetic fields. A detailed investigation
of the damping effect was done in ref.~\cite{calzetta01} in the model
with $N$ charged scalar fields in large $N$ limit. The conclusion of
the work was that the magnetic field may exponentially rise and though
with the parameters of the model the resulting field was too weak,
a much higher intensity is still not excluded.

A breaking of conformal invariance can be achieved 
by the condensate $\langle \phi^\dagger \phi \rangle$ which is formed 
due to infrared instability of light scalars in de Sitter 
space-time~\cite{infrared}. A possibility to generate primordial 
magnetic field through this phenomena was explored in 
ref.~\cite{davis01}. Because of the coupling 
\bear
\left[ \left( \partial_\mu -ie A_\mu \right) \phi \right]^\dagger
\left[ \left( \partial^\mu -ie A^\mu \right) \phi \right] \rar
e^2 | \phi |^2 A^\mu A_\mu
\label{e2phia2}
\ear
this condensate effectively gives a non-zero mass to photon that
violates conformal invariance. In the subsequent course of cosmological
expansion the condensate of $|\phi|^2$ would disappear and the
sudden vanishing of the effective photon mass could create sufficiently
strong magnetic fields to seed galactic dynamo.

An interesting scenario is suggested in ref.~\cite{dimopoulos01}, 
which is a modification of the one described above. 
It is based on the breaking of conformal invariance by the Z-boson 
field due to the coupling of $Z$ to the electroweak Higgs field. At 
the end of inflation reheating restores EW symmetry and the Z field 
is transformed into weak hypercharge field and at the EW phase 
transition the latter turns into magnetic field. The process could
be efficient enough to seed the observed galactic fields.
(The quoted paper contains also  an extensive list or references.)

A study of parametric resonance amplification of the generation 
of magnetic fields during preheating and inflation was performed
in ref.~\cite{bassett01} for several different models of breaking
of conformal invariance. A possible relevance of parametric resonance
to the problem of primordial magnetic field was noticed in the
paper~\cite{finelli01}.

Another idea to break conformal invariance and as a result to amplify
vector fluctuations during inflation was suggested in the
paper~\cite{gasperini00a}. In supergravity theories the photon field
can be mixed with the so called graviphoton, i.e. the massive vector 
component $V_\mu$ of the gravitational supermultiplet:
\bear
{\cal L} = - F_{\mu\nu} F^{\mu\nu}/4  -  G_{\mu\nu} G^{\mu\nu}/4
 +\zeta F_{\mu\nu} G^{\mu\nu} +m^2 V_\mu V^\mu,
\label{AV}
\ear
where $G_{\mu\nu} = \partial_\mu V_\nu -\partial_\nu V_\mu$,
and $\zeta$ is a dimensionless constant. Earlier a model of two-photon
mixing was proposed in the paper~\cite{okun82} in a different connection.

All models mentioned in this section could easily give large scale fields 
because they are operative at inflationary stage and the characteristic
wave length of the field could be exponentially huge, but the amplitude 
of the field is usually smaller than the necessary value so that galactic 
dynamo~\cite{zeldovich83} should be invoked to amplify these seed field 
by more than 10 orders of magnitude. However, the dynamo may have problems
in producing large scale magnetic fields because of fast saturation,
see e.g. the review~\cite{kulsrud97}. Moreover, it seems impossible 
to explain strong magnetic fields observed in clusters of galaxies
by the dynamo amplification~\cite{colgate00}.

\section{Cosmological Phase Transitions.}

The idea that phase transitions in the early universe might have produced 
seed magnetic fields was pioneered in ref.~\cite{hogan83}. Expanding
bubble walls between two phases could generate electric currents which in
turn produce magnetic fields. Bubble collisions or hydrodynamical
instability on the bubble walls could create turbulence flows and give
rise to magneto-hydrodynamic (MHD) amplification of the magnetic fields.
Such models were studied in series of papers both for electro-weak 
(EW)~\cite{vachaspati91b} and QCD phase transitions~\cite{olinto94}
(see more references in the review~\cite{grasso01}). However the existing 
experimental data indicating that the Higgs boson is heavy makes first 
order electro-weak phase transition rather improbable and there are 
also some doubts if the QCD phase transition is first order. 
Furthermore, even if phase transitions are first order, still in such 
a scenario is hardly 
possible to create large scale magnetic fields. Indeed, the comoving
coherence length of the magnetic field cannot be larger than the Hubble
horizon at the phase transition, which is much smaller than the galactic
size. Though the coherence length may grow due to MHD effects, this 
happens at the expense of the magnetic field strength. According to 
ref.~\cite{son99} neither EW or QCD phase transition could create 
sufficiently large magnetic fields on the galactic scales.

An attempt to overcome the large scale problem was done in the 
papers~\cite{iwasaki97} under assumption that there could exist
cosmological domain walls and that the fermions living on the
wall could develop spontaneous magnetization orthogonal to the wall. 
This idea was further pursued in the recent works~\cite{forbes00}. 
However, as was argued in refs.~\cite{voloshin96}, magnetization is 
either absent or, even if it is non-zero, a magnetized domain wall
cannot produce a correlated on large scale cosmologically interesting
magnetic field.

More efficient could be generation of magnetic fields by cosmic, 
possibly superconducting, strings~\cite{vachaspati91,vachaspati92}.
String motion would create large scale vorticity perturbations inside 
their wakes and according to ref.~\cite{harrison69} vorticity should 
produce electric currents and magnetic fields. According to
ref.~\cite{dimopoulos98}, superconduction strings could produce 
magnetic fields even more effectively because they might carry very
large electric currents generated by their motion through cosmological 
plasma. On the other hand, this scenario is strongly restricted
by the recent measurements of the angular fluctuation of cosmic
microwave background radiation (see e.g. the reviews~\cite{cmbr}) 
that put very restrictive limits on the cosmological density of cosmic 
strings.

\section{Chaotic Electric Currents at Large Scales.}

Chaotic electric currents could be generated in the universe if an 
inhomogeneous charge asymmetry between particles and antiparticles
had been created at an earlier stage. When the wave length of the 
inhomogeneity becomes smaller than the horizon, some turbulent flows
would evolve and give rise to non-vanishing electric currents. 
As is argued in ref.~\cite{dolgov-s93}, if at an early stage of the 
evolution of the Universe the gauge invariance of electromagnetism 
was spontaneously broken, an electric charge asymmetry would develop 
through the same mechanism as cosmological baryon asymmetry. Electric
asymmetry could be inhomogeneous if the mechanisms similar to
creation of isocurvature perturbation in the baryon sector were
operative (see e.g. the review~\cite{dolgov92}). Gauge
invariance must be restored at low temperatures and after its
restoration the asymmetry should disappear so that the net electric 
charge density must vanish. The compensating charge could be
produced from the Higgs vacuum in the form of heavy charged particles.
Since the electric asymmetry was inhomogeneous, the number density of
these particles would be inhomogeneous as well. Correspondingly energetic 
products of their decay would create an electric current and 
a local charge asymmetry. 

Alternatively, such an asymmetry could be created even if the electric 
current was always conserved but a cosmological asymmetry in another 
nonconserved charge existed. The characteristic length of the inhomogeneities
could be astronomically large if the conditions for their creation were
cooked during inflation. The primary currents which created the asymmetry 
as well as those damping it via plasma discharge could generate chaotic 
magnetic fields on astronomically interesting scales. These fields might 
be large enough to seed the observed magnetic fields in galaxies via a 
protogalactic dynamo.

Another model of similar type was recently suggested~\cite{dolgov01},
which practically does not demand any new physics and, in contrast to 
all discussed above models, operates at very low energies in MeV range.
In the simplest version of the scenario the only necessary assumption is
that of light sterile neutrino, $\nu_s$, weakly mixed with the ordinary 
active ones, i.e. $\nu_e$, $\nu_\mu$, or $\nu_\tau$. Neutrino oscillations 
in the early universe could create a very large lepton asymmetry in the 
active neutrino sector, 
$(n_\nu - \bar n_\nu) / (n_\nu + \bar n_\nu) > 0.1$
if MSW resonance condition is fulfilled, i.e. if sterile neutrinos are
lighter than active ones~\cite{foot96} (more references where this effect
is discussed can be found in recent papers~\cite{foot00}). Moreover,
as was found in ref.~\cite{dibari00} this large lepton asymmetry should 
be strongly inhomogeneous on superhorizon scale, if there were very small
inhomogeneities in the primordial lepton or baryon asymmetries. Another
possible mechanism~\cite{dolgov91,dolgov92} of generation of inhomogeneous 
and large lepton asymmetry is based on Afleck and Dine scenario of 
lepto/baryogenesis~\cite{affleck85}.

If inhomogeneous cosmological lepton asymmetry was indeed generated by
one or other mechanism mentioned above, neutrino currents should be
developed along the density gradients  when the neutrino mean free path 
$\ell_\nu(T)$ grew and became comparable to the wave length of the
inhomogeneity, $\lambda$. Elastic scattering of the diffusing neutrinos 
on electrons and positrons in the primeval plasma would be able to 
accelerate the electron-photon fluid producing vorticity in the plasma.
Depending on the amplitude and wavelength of the fluctuations of the 
charge asymmetric difference,
$n_{\nu_a}({\bf x}) - n_{\bar\nu_a}({\bf x})$
($a = e,\mu,\tau$),
this period could be sufficient for the MHD engine to generate magnetic
field in equipartition with the turbulent kinetic energy. The seed
field required to initiate the process arises naturally as a consequence 
of the difference between the $\nu_a e^-$ and $\nu_a e^+$ cross sections 
and of the neutrino-antineutrino local asymmetry. 

Starting form the Boltzmann equation one can obtain the equation
describing evolution of the average flux of $i$-th component of
neutrino momentum
\begin{equation}
\frac{\partial}{\partial t} {K}_i({\bf x},t)+4 H\,K_i({\bf x},t)
+\frac{\partial}{\partial x^j} {K}_{ij}({\bf x},t)
= -\tau_w^{-1} K_i
\label{diffeq}
\end{equation}
where 
\bear
{K}_i = \int k_i~f_\nu (E,{\bf k}) \frac{d^3{\bf k}}{(2\pi)^3},
\label{ki} ~~~~ {K}_{ij} = \int {k_i k_j \over E}~f_{\nu}(E,{\bf k}) 
\frac{d^3{\bf k}}{(2\pi)^3} \nonumber
\ear
In the above $E$ and ${\bf k}$ are respectively the neutrino energy and
spatial momentum, $H$ is the universe expansion rate, $\tau_w$ is the 
effective weak interaction time, and  $f_{\nu}(E,{\bf k},t, {\bf x})$ is
the neutrino distribution function.
The source of magnetic field is proportional to curl of electric current,
${\bf \nabla}\times {\bf J}$, which in turn is proportional to the local 
vorticity of the source term $\partial_j K_{ij}$ in kinetic 
equations~(\ref{diffeq}). The latter is nonvanishing  
for an anisotropic random initial distribution of neutrino leptonic
charge and is numerically close to $\partial_j K_{ij}$ 
divided by $\lambda$.

Using equation of motion of electron-positron fluid disturbed by the 
neutrino flux one can estimate the magnitude of electric current induced
by neutrinos as
\bear
J_{\rm ext} = 4\cdot 10^{-20}e T^3 \left({T\over {\rm MeV}}\right)^3
\left({\delta n_\nu \over n_\nu}\right)_\lambda~, 
\label{Jext}
\ear 
which creates the seed field with the strength 
$B^{\rm seed}_\lambda \approx 10^{-22}  
\left( T / {\rm MeV}\right)^2$ at the time  $t/\lambda \sim 1$. In the
equation above $e$ is the charge of electron, $T$ is the plasma 
temperature, and $\delta n_\nu$ is the magnitude of the fluctuation
in neutrino number density. 

The evolution of magnetic field is given by the equation
\bear
\partial_t {\bf B} + 2H{\bf B} = 
{\bf \nabla} \times \left( {\bf v}\times {\bf B}\right)
+ \gamma^{-1}  {\bf  \nabla} \times {\bf J_{ext}}
\label{dtB}
\ear
where $\gamma$ is (large) electric conductivity of relativistic 
cosmological plasma. Numerical solution of this equation~\cite{dolgov01}
showed that magnetic field quickly enough approaches energy equipartition
with the fluid motion. 

Such mechanism cold give rise to magnetic fields of intensity 
$B_0 \sim \times 10^{-5}$ Gauss at the present time with a coherence 
length $\lambda_0 = \lambda_d~ r_H(t_d)~ (T_d/T_0) \simeq 10^2
~\lambda_d$ pc. Galactic magnetic fields are observed with characteristic
strength of the order of $1~\mu G$ extending over scales $\sim 1$
kpc \cite{kronberg94}. Taking into account flux conservation during
the protogalaxy collapse, the primordial origin of galactic fields
would require a protogalactic field with the strength $\sim 10^{-10}$
Gauss and the coherence length of 0.1 Mpc \cite{kronberg94,grasso01}.
Although this scale is much larger than the coherence length predicted 
by the discussed model, it is natural to expect that some
homogenization could take place during galaxy formation. Since the
field orientation is random over scales larger than $\lambda_0$,
the predicted mean field on the protogalactic scale will be
obtained by a suitable volume average~\cite{hogan83}
\bear
B(0.1~{\rm Mpc}) \simeq B_0 \left(\frac {\lambda_0}{0.1~{\rm
Mpc}}\right)^{3/2} \simeq 10^{-10}b~{\lambda_d}^{3/2}~{\rm G}.
\label{bnow}
\ear
One can conclude from this expression that galactic magnetic fields
may be produced by the neutrino number fluctuations with the relative
amplitude $\sim 1$ extending over scales comparable to the Hubble horizon 
at neutrino decoupling~\cite{dolgov01}.

\section{Conclusion}\label{sec:concl}

Thus, it seems that the models that invoke inflation for creation
of large scale magnetic fields are in a better shape than others.
Possibly the existence of galactic magnetic field might be considered 
as an additional indication to inflation. On the other hand, all the
concrete scenarios based on inflationary stretching of the magnetic 
fields, generated in the early universe, are heavily based on new
physics, they demand an introduction of new fields or interactions
and their predictive power is rather poor, simply because there is
a plethora of the models and it is difficult to judge which one is
the real, without knowledge the physics at very high energies far
beyond the reach of the present day accelerators.

The scenarios using non-equilibrium phase transitions or
topological defects (especially domain walls) encounter serious 
difficulties and at the present day look as outsiders. Moreover, 
astronomical data are rather against abundant cosmic topological 
defects and particle physics indicates that EW and QCD phase 
transitions are most probably second order. The conclusion that 
the mechanisms based on phase transitions are unlikely is supported
by an astrophysical analysis performed in ref.~\cite{battaner00}.

It would be very nice if a mechanism of creation of seed magnetic
fields at a later stage of cosmological evolution is found. It would
make a problem of the coherence length less severe because of a larger
Hubble horizon and the underlying physics could be accessible to 
direct tests. The nearest to this request is the model of
ref.~\cite{dolgov01}, according to which magnetic field is generated
at temperatures of a fraction of MeV and the only essential assumption
of the model is an existence of light sterile neutrino weakly mixed 
with one or several usual neutrino flavors. Still the problem of
the large scale is present in this model as well, though at a much
weaker level. Possibly the most efficient mechanism would be based on 
the assumption of a new light and very long-lived particles whose decays 
could induce electric current on astronomically big scales and in turn 
generate the seed magnetic field. However it is difficult (if possible) 
to satisfy existing cosmological and astrophysical constraints on the
properties of such particles. 

\section*{Acknowledgments}

I am grateful to D. Grasso and T. Prokopec for helpful comments.

\end{document}